# Optical focusing inside scattering media with time-reversed ultrasound microbubble encoded (TRUME) light

Haowen Ruan[1,†,*], Mooseok Jang[1,†,*], Changhuei Yang[1]

[1]*Electrical Engineering, California Institute of Technology, 1200 East California Boulevard, Pasadena, California, 91125, USA*

[†]These authors contributed equally to the work.

[*]Correspondence and requests for materials should be addressed to H. R. (hruan@caltech.edu) or M. J. (mjang@caltech.edu).

Focusing light inside scattering media in a freely addressable fashion is challenging, as the wavefront of the scattered light is highly disordered. Recently developed ultrasound-guided wavefront shaping methods are addressing this challenge, albeit with relatively low modulation efficiency and resolution limitations. In this paper, we present a new technique, time-reversed ultrasound microbubble encoded (TRUME) optical focusing, which is able to focus light with improved efficiency and sub-ultrasound wavelength resolution. This method ultrasonically destructs microbubbles, and measures the wavefront change to compute and render a suitable time-reversed wavefront solution for focusing. We demonstrate that the TRUME technique can create an optical focus at the site of bubble destruction with a size of ~2 μm. Due to the nonlinear pressure-to-destruction response, the TRUME technique can break the addressable focus resolution barrier imposed by the ultrasound focus. We experimentally demonstrate a 2-fold addressable focus resolution improvement in a microbubble aggregate target.

Creating an optical focus inside a scattering medium, such as biological tissue, has great potential in various applications. However, optical scattering, as a dominant light matter interaction within biological tissue, poses a very significant challenge. Recent developed wavefront shaping techniques have begun to address this[1–4] by exploiting the deterministic and time-symmetric nature of scattering. Focusing light through scattering media has been realized by iterative optimization methods[2,5], optical phase conjugation (OPC)[6,7], and direct measurement of the transmission matrix at large scale[8–10].

Determining the correct wavefront to focus light from outside of a scattering medium to a point within requires a feedback or tagging mechanism. Typically, these mechanisms arise from a localized "guidestar" point. Examples of guidestars include second harmonic generation[11], fluorescence[12,13] and kinetic[14,15] targets. While individual guidestars enable light shaped to focus to their physical location, these techniques fundamentally lack addressability if dense and randomly distributed sets of guidestars are present.

Alternatively, ultrasound-assisted techniques, such as photoacoustic-guided[10,16–18] and time-reversed ultrasonically-encoded (TRUE)[19–21] optical focusing techniques, employ a focused ultrasound beam as a "virtual guidestar". Unlike the above techniques, it is easy to move or scan an ultrasound focus to new positions. While TRUE has a speed advantage over the photoacoustic approach, TRUE guidestar is generally weak. Typically <1% of the probe light field that passes through the ultrasound focus is tagged[22,23]. Moreover, the resolution achieved is limited by the ultrasound focus size. Although more advanced TRUE techniques, such as iterative TRUE (iTRUE)[24–26] and time reversal of variance-encoded light (TROVE)[27], may break this resolution barrier, it comes at the expense of time. For practical biological applications with tight time constraints, efficient and fast techniques are highly desired.

Here, we present a high resolution, deep tissue optical focusing technique termed time-reversed ultrasound microbubble encoded (TRUME) optical focusing. Microbubbles have been widely used in ultrasonic imaging as ultrasound contrast agents because they generate stronger echoes and nonlinear acoustic signals as compared with surrounding tissue[28,29]. They are also helpful for ultrasound modulated optical imaging inside scattering media[30–32]. Furthermore, like fluorescent markers, microbubbles can be modified to bind to selected biomarkers, suggesting promising applications in functional imaging and therapeutic applications[28].

We demonstrate that the selective nonlinear destruction of microbubbles with a focused ultrasound beam can serve as effective, highly localized and freely-addressable guidestar mechanism. In brief, TRUME works by measuring the scattered optical field before and after the ultrasonic destruction of the microbubble. Subsequently, by playing back the phase conjugate of the difference of these two fields, TRUME can generate a focus at the location of the destructed microbubble. Although multiple foci could be created at the same time when multiple microbubbles are present within the original ultrasound focus, we show that careful selection of the ultrasound pressure can lead to destruction of microbubbles in an addressable volume that is smaller than the ultrasound focus. This is a result of a nonlinear pressure-to-destruction response curve associated with the microbubbles. This technique combines the advantages of both physical and virtual guidestars to provide an efficient, fast and addressable deep tissue optical focus.

## Results

### Principles

Our TRUME setup uses a digital optical phase conjugation (DOPC) system as its wavefront recording and playback engine[7,25] (Figure 1a). In the recording phase, the scattered field from the sample is recorded by the camera of the DOPC system. In the playback step, a suitable pattern is displayed on the spatial light modulator (SLM) and a collimated 'blank' beam is modulated to form the playback light field. Precise alignment of the camera and SLM allows high fidelity phase conjugate playback of the record field. Experimentally, this DOPC system is able to control ~$10^5$ optical modes simultaneously[33].

Here, we demonstrate TRUME in transmission geometry (Figure 1a), in which a sample beam transmits through the sample in the z direction and part of the scattered light is measured by the camera on the other side of the sample. An ultrasound beam is focused on the microbubbles embedded between two diffusers through water coupling. TRUME operates in three steps. First, an optical field (Field A) is measured by the camera (Figure 1b) with phase shifting digital holography[34]. Second (Figure 1c), ultrasound is applied to destruct the targeted microbubble, immediately followed by the measurement of a second optical field (Field B). The difference of the fields (Field A – Field B) is the scattered field solution associated with the microbubble. The DOPC system computes this difference field and plays back a phase conjugate copy. Since the difference field primarily contains information from the microbubble only, the conjugated beam focuses to the position of the destructed microbubble (Figure 1d).

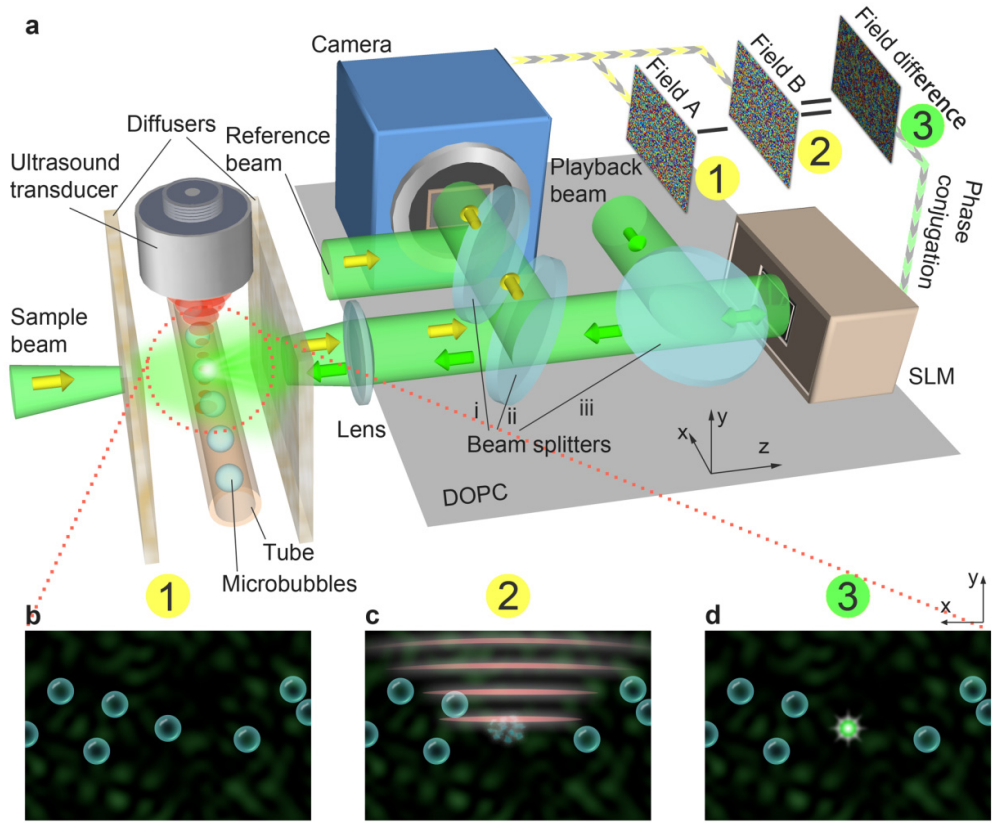

**Figure 1 | Principle of TRUME technique.** (a) Illustration of the experimental setup. The microbubbles perfuse inside an acrylic tube, which is sandwiched between two diffusers. A DOPC system is used as a phase conjugation mirror to time-reverse the light back to the sample. (b)-(d) Illustration of TRUME optical focusing technique in 3 steps. At the first step, the camera of the DOPC system captures a transmitted optical field (Field A) before applying ultrasound to the sample (b). Ultrasound bursts are then used to destruct the targeted microbubble (c), resulting in a different optical field (Field B). The difference between two fields yields an optical field that appears to emerge from the destructed microbubble. The conjugated phase of the difference field is then sent to the SLM to create a playback beam, which focuses light at the position of microbubble destruction (d).

TRUME shares the same mathematical framework as guidestar techniques using kinetic objects[14,15]. The optical field on the target plane $E_t$ can be decomposed into a microbubble diffracted field $E_m$ and a background field $E_b$, which describes the field after microbubble destruction: $E_t = E_m + E_b$. Since the camera and SLM contain discrete components, it is convenient to discretise $E_m$ and $E_b$ as column vectors with $n$ complex elements, with each element mapping to an optical mode on the two-dimensional target plane. We may then connect this target field to the field on the measurement plane $E_t'$ through a matrix equation: $E_t' = TE_t = T(E_m + E_b)$. Here $T$ is an $m \times n$ matrix describing the scattering medium and $E_t'$ is a column vector of $m$ elements, with each element mapping to an optical mode on the two-dimensional measurement plane. Similarly, the field measured after microbubble destruction can be given by $E_b' = TE_b$. The difference field on the measurement plane is thus,

$$\begin{aligned} E_d' &= E_t' - E_b' \\ &= T(E_m + E_b) - TE_b \\ &= TE_m \end{aligned} \quad (1)$$

Here, subtraction effectively removes the impact of the background field on the measurement plane, resulting in a field that appears to be scattered from the microbubbles only. Finally, we play back the conjugated field $E_d'^*$ with an optical gain $\alpha$ provided by the playback beam (Figure 1a). Assuming time-reversal symmetry, we may express playback as a multiplication with $T$ from the left with the conjugate transpose of the difference field. Therefore, the playback field $E_p$ on the target plane takes the form:

$$\begin{aligned} E_p &= \alpha E_d'^* T = \alpha (TE_m)^* T \\ &= \alpha E_m^* T^* T \approx \alpha \beta E_m^* \end{aligned} \quad (2)$$

Here, we assume minimal absorption within the sample to apply the approximation $T^*T \approx \beta I$, in which $\beta$ is the fraction of scattered light field that is measured in the DOPC system and $I$ is an identity matrix. The playback light effectively cancels out the random transmission matrix to refocus at the location of microbubble destruction.

**Visualization and efficiency characterization of the focus**

To validate TRUME focusing, we directly visualized the target plane using a 10X microscope system (see Methods) before and after the TRUME procedure. In this experiment, we shifted the front diffuser along the x direction (to the "open position" in Figure 2a) for direct imaging of the target plane during the focusing phase. The target sample here is microbubbles embedded in agarose gel within an acrylic capillary tube (see Methods) as shown in Figure 2b. Immediately after measuring the first optical field, a 20 MHz focused ultrasound beam was used to destruct one microbubble, followed by the measurement of the second field. We then image the target plane again to confirm the destruction of the microbubble (Figure 2c) and directly visualized the focus created at the position of destructed microbubble (Figure 2d). The measured peak intensity to background intensity ratio (PBR) of the TRUME focus in Figure 2d is ~510.

For comparison, we also measured the focusing profile of TRUE (Figure 2e). The PBR of the TRUME focus is around two orders of magnitude higher than that of TRUE (PBR = ~2 in Figure 2e), since the TRUME concentrate light at fewer optical modes and has a stronger modulation efficiency per mode.

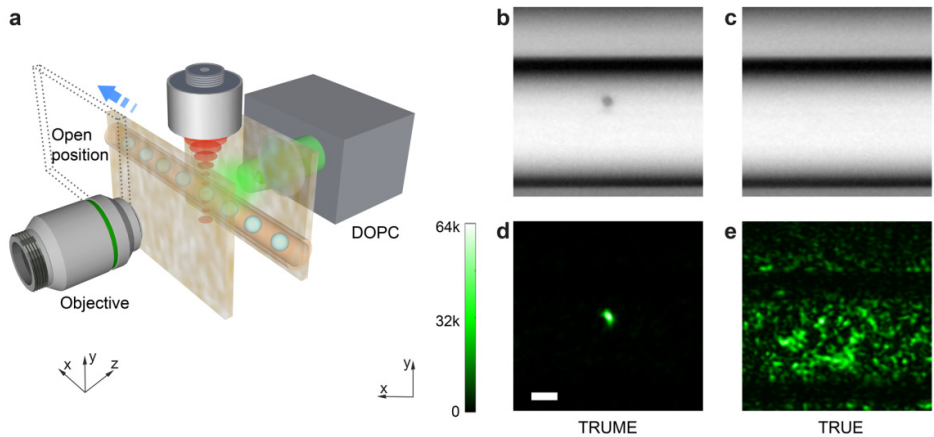

**Figure 2 | Visualization of the target plane.** (a) Illustration of the observation setup. The front diffuser was shifted to the open position before and after TRUME for direct visualization. A 10X microscope system was used to observe the target plane. (b), (c) Images of a microbubble before and after applying ultrasound. (d) Optical focus created at the position of microbubble destruction. (e) Focusing results of TRUE technique. Scale bar: 10 um.

We separately measured the modulation efficiency of ultrasound in a clear medium, and found that ~0.5% of light passing through the ultrasound focus (2 MPa peak pressure) is modulated. In comparison, the proportion of light passing through the location of the bubble that is modulated by bubble destruction reaches ~25%. This large difference in modulation efficiency is the primary reason why the TRUME guidestar offers a stronger focus.

**Deep tissue optical focusing**

To study the performance of TRUME for focusing through biological tissue, we used two pieces of 2-mm thick biological tissue as our diffusive medium (see Methods). The experimental setup matches that shown in Figure 2a. The images of the microbubble before and after destruction are shown in Figure 3a and b. An optical focus (Figure 3c, e) was created using TRUME, with PBR of ~23. Fitted Gaussian profile (to the one-dimensional data through the centre of the focus in the x and y directions) shows the focus full width at half maximum (FWHM) (Figure 3e) is 2.4±0.2 μm in the x direction and 1.7±0.2 μm in the y direction (95% confidence bound). To confirm that this optical focus was created due to optical phase conjugation, we shifted the SLM phase pattern in both x and y directions by 10 pixels. As shown in Figure 3d, the optical focus vanishes, as expected. The optical fields measured before and after microbubble destruction, as well as the subtracted field, are shown in Figure 3f-h, respectively.

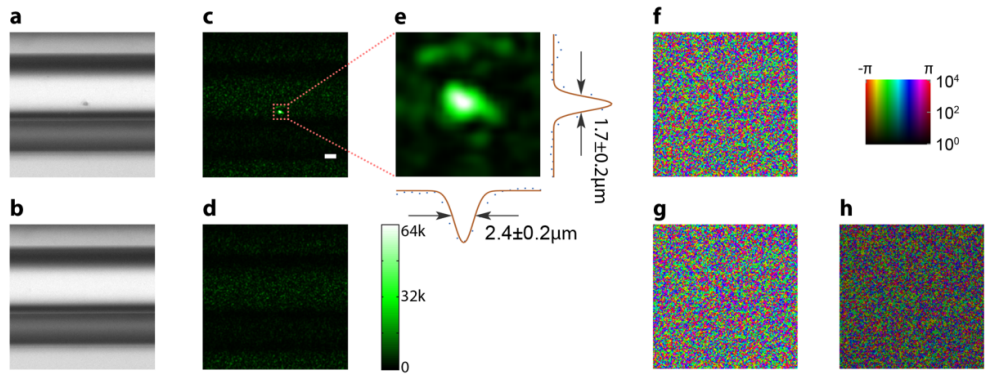

**Figure 3 | Optical focusing in 2-mm deep chicken tissue.** Two pieces of 2-mm thick chicken tissues were used as diffusers. (a) A microbubble in a tube before destruction. (b) After destruction. (c) A light focus was created at the position of the destructed microbubble (PBR ~23). (d) The optical focus vanished as the SLM shifts 10 pixels in both x and y directions. (e) 10X zoom-in image of the optical focus with quantified resolution. (f), (g) Central part (200 pixels by 200 pixels) of the optical fields captured before (f) and after (g) the destruction of the microbubble. (h) Difference of the fields in (f) and (g). Scale bar: 10 µm.

**Demonstration of flow stream monitoring**

TRUME may help perform cytometry behind a scattering media. Microbubbles are currently used as contrast agents in blood circulation ultrasound imaging[28]. To demonstrate this potential application (Figure 4a), we mixed fluorescent microspheres (4 µm) and microbubbles in 1X phosphate buffered saline (PBS) and pumped the solution through an acrylic tube (see Methods). We first formed an optical focus, as shown in Figure 4b, by implementing TRUME on a microbubble at the target location. Fluorophores that subsequently flow across the focus then interact with the focused light spot to emit fluorescence. The fluorescence was filtered with an emission filter and detected by a single photon counting module (SPCM) outside the scattering medium (see Methods). The resulting signal is shown in Figure 4c. After counting, the front diffuser was shifted to the open position and the fluorescent microspheres were imaged with an emission filter for verification (Figure 4d). The agreement of the results positively validates this proof-of-concept.

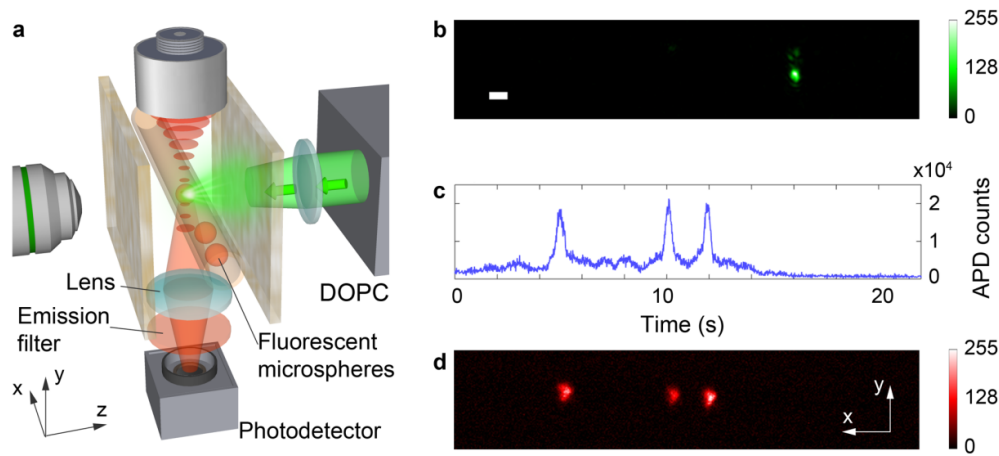

**Figure 4 | Demonstration of flow stream monitoring through a scattering sample.** (a) Illustration of the experimental setup. An external SPCM was used to detect the excited fluorescence through the fluorescence filter. (b) A light focus was created with TRUME. (c) Photon counts recorded by the SPCM as the optical focus probed the flowing microspheres. (d) Image of the fluorescent microspheres after passing through the optical focus in the x direction. Scale bar: 10 μm.

## Addressable focus resolution improvement with nonlinear microbubble destruction.

Our demonstrations of TRUME thus far destruct an isolated microbubble with a relatively large ultrasound focus (one to two orders of magnitude larger), forming one sharp optical focus. If multiple microbubbles are clustered together, then the ultrasound focus may destroy more than one bubble. In this scenario, TRUME will generate an optical "focus" that can be significantly broader than the focus we have discussed thus far. To distinguish the two focus types, we will use the term addressable focus to refer to the achievable TRUME focus in the scenario where microbubbles are dense.

The addressable focus size is statistically determined by the pressure-to-destruction response of the bubbles. Interestingly, the probability of microbubble destruction varies nonlinearly as a function of pressure. In the ideal case where all microbubbles have the same destruction threshold, one can set the peak ultrasound pressure to be right at the threshold so that only the microbubble at the centre of the ultrasound focus can be destructed and obtain addressable focus size that is equal to the single bubble TRUME focus size. In practice, the actual pressure-to-destruction response curve is not a simple step function. Nevertheless, the more nonlinear the response curve is, the sharper addressable focus we can achieve with TRUME.

To better characterize the pressure-to-destruction response and determine the TRUME addressable focus resolution achievable with our system, we experimentally measured the cumulative distribution function of the microbubble destruction $\sigma(P)$ by counting the number of microbubble destroyed as a function of pressure. As shown in Figure 5a (red), the cumulative distribution function reveals a strong nonlinear relationship between destruction probability and pressure. Given a focused ultrasound profile $P(x)$ (Figure 5a, green, see also Methods), we were able to calculate the microbubble destruction probability over positon $\sigma(P(x))$ (Figure 5a, blue), which predicts the addressable focus resolution of TRUME. The resulting profile is significantly narrower than the ultrasound pressure profile, implying that the nonlinear relationship would effectively improve the addressable focus resolution of TRUME.

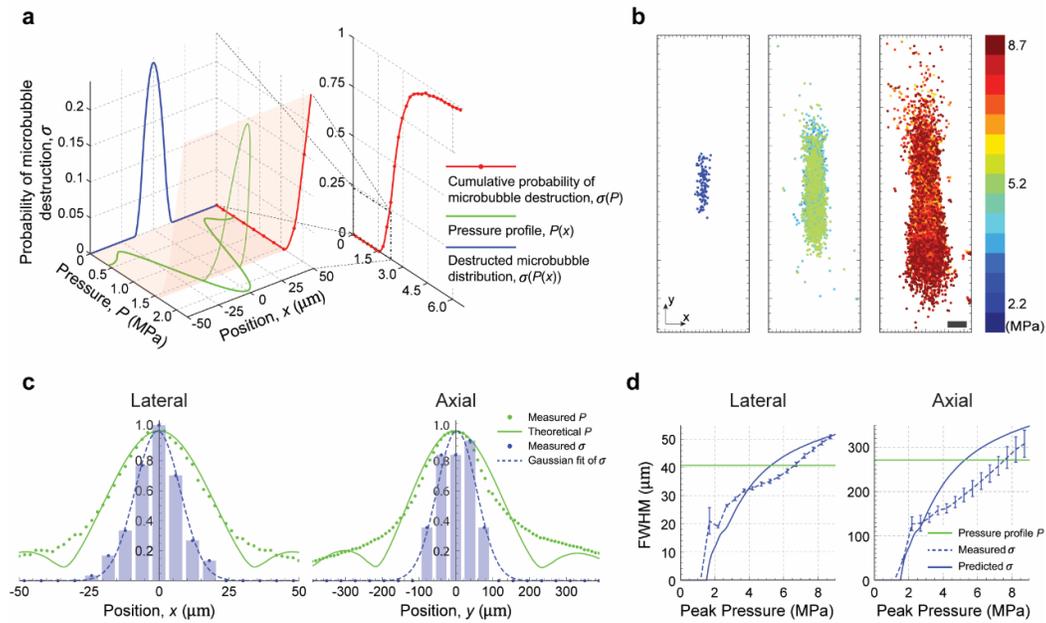

**Figure 5 | Addressable focus resolution improvement by exploiting nonlinear microbubble destruction.** (a) Calculation of microbubble destruction probability distribution over position (blue) based on the measured cumulative distribution function of the microbubble destruction $\sigma(P)$ (red) and the theoretical ultrasound pressure profile $P(x)$ (green). (b) Two-dimensional distribution map of TRUME foci over pressure levels. (c) Comparison of the TRUME focus probability distribution (histograms, with Gaussian fit) and the ultrasound pressure profile (green) in both lateral (left) and axial (right) directions. The histograms were calculated from the low-pressure map (left figure in b). (d) Theoretical (i.e. FWHM of the blue curve in a) and experimental (i.e. FWHM of the blue curve in c) TRUME addressable focus resolution over pressure. Green lines mark the FWHM of the ultrasound profile. Error bar indicates 95% confidence bound. Scale bar in b: 50 μm.

To experimentally confirm the improvement of addressable focus resolution of TRUME, we used a thin microbubble sheet (see Methods) to visualize the distribution of the foci as ultrasound pressure increases. In order to cover the entire ultrasound focus (-6 dB) with the current observation system and further improve the resolution, we used a 45 MHz, high numeric-aperture ultrasound beam with a measured beam diameter of ~40 μm and focal zone of ~270 μm (-6 dB) (see Methods) in this experiment. We applied 15 levels of ultrasound pressure (linearly from 1.7 to 8.7 MPa) to the sample and measured the fields before and after each insonation. We then played back the corresponding field difference sequentially, recorded the resulting focus patterns, and applied a watershed algorithm to extract each focus centroid (see Methods). To collect meaningful statistics, this process was repeated 135 times at different unaffected regions of the microbubble sheet. We aggregated the measured TRUME focus centroids into a statistical map as shown in Figure 5b, where foci are displayed in three pressure groups. The profile of the foci broadens as the ultrasound pressure becomes higher, confirming the nonlinearity effect in TRUME.

To quantify the addressable focus resolution improvement, we calculated the FWHM of Gaussian profiles that are fitted to the histograms of each statistical map along both lateral (x) and axial (y) directions. Figure 5c shows the Gaussian fits and histograms of the lower pressure group (<2.2 MPa, Figure 5b, left) where microbubbles start to collapse. We also measured the ultrasound pressure profiles, which closely matches with its theoretical profile in both directions (see Methods). The FWHM of the Gaussian fit to the centroid histogram in the lateral (x) direction is 19 μm, while that of theoretical ultrasound focus is 40 μm. Likewise, the FWHM of TRUME addressable focus along axial (y) direction is 130 μm, which is also lower than the ultrasound focus (270 μm). We further studied the effect of ultrasound pressure on the nonlinearity induced resolution improvement by calculating the FWHMs of the Gaussian fits of both theoretical microbubble destruction distribution (e.g. blue curve in a) and TRUME focus histogram profiles (e.g. blue curve in c) at various pressure levels. As shown in Figure 5d, both experimental and theoretical FWHMs are lower than that defined by the ultrasound focus (green line) when the ultrasound pressure is less than ~5 MPa. The discrepancy between these two curves is attributable to variations of the samples.

## Discussion

Combining the advantages of a physical and a virtual guidestar, TRUME can selectively focus light to a size of ~2 μm in deep tissue, given the distribution of microbubbles it targets is sufficiently sparse. When the microbubble distribution is dense, we show that TRUME may still achieve an addressable focus resolution ~2X higher than that defined by its ultrasound focus. As this method simply requires two measurements and no iterations, it is intrinsically fast and a good match with *in-vivo* applications. Next, we list several factors that affect TRUME performance, and outline several of its potential applications.

The size of an individual focus depends on that of the microbubble which is typically at micrometre scale, ~10-fold smaller than a TRUE focus. Although ultrasound focus could cover multiple microbubbles, TRUME further confines the targeting range by taking the advantage of the nonlinear relationship between destruction population and ultrasound pressure. The addressable focus resolution improvement was largely limited by the broad size distribution of the microbubbles, and thus can be enhanced by reducing the standard deviation of the radius of microbubbles, via separation techniques[35] or methods based on established protocols[36,37]. Alternatively, simultaneously focusing to multiple microbubble locations might also be a desired experimental goal, like when using microbubbles as selective markers (e.g. binding to certain disease markers)[28].

The PBR of TRUME is measured to be ~100-fold higher than that of TRUE (~510 versus ~2, using a ground glass diffuser sample and the setup in Figure 2). Two factors lead to this large PBR increase. First, TRUME practically encodes significantly fewer optical modes, even if multiple microbubbles are present within the ultrasound focus. Second, the modulation efficiency of TRUME is much higher than TRUE. In our experiment, we found ~25% modulation of the light passing through the TRUME guidestar. In comparison, a TRUE guidestar with a peak pressure of 2 MPa only modulates ~0.5% of its contained light.

Taking the advantage of parallel field measurement, this DOPC-based technique creates optical foci in hundreds of milliseconds (~280 ms in our experiments), a timescale short enough for *ex-vivo* or even some *in-vivo* biological applications[38]. It should be noted that no frame averaging was needed for any of our TRUME experiments. Its operation speed is limited by the DOPC system frame rate, which can be improved with an FPGA based system. Off-axis holography based field measurement or binary phase measurement would further improve the system speed by reducing the number of frames needed for field measurement.

The time needed to destruct a microbubble depends on the mechanisms of microbubble destruction, which can be classified into fragmentation and diffusion[39]. Fragmentation occurs when ultrasound pressure is relatively high and the microbubble is destroyed within the timescale of microseconds, which is ideal for TRUME in terms of operation speed. However, if low ultrasound pressure is used, acoustic driven diffusion is the dominant destruction mechanism. This process typically spans tens of microsecond, depending upon both the ultrasound parameters (pressure, frequency, cycles, etc.) and microbubbles properties (size, shell material and encapsulated gas)[39]. In this paper, the ultrasound duration was 28.6 ms (one camera frame period), within which incomplete gas dissolution was also observed under certain circumstances, such as with low ultrasound pressure and a large microbubble diameter. This effect results in a size decrease rather than complete disappearance of the microbubble. Intriguingly, decreasing the size of the microbubble between capturing two optical fields also enables TRUME to form an optical focus at the targeted microbubble, because it shares the same effect as the complete microbubble destruction – inducing difference between two optical fields.

Microbubbles are usually made with albumin or lipid, which stabilizes high molecular weight gases, such as perflutren. These microbubbles have been widely used as ultrasound contrast agents and proven for some applications in the human body. Their biocompatibility makes them a promising optical guidestar in biological tissue. Besides ultrasonic imaging, microbubbles also have promising applications in gene and drug delivery[40], where their ultrasonic destruction can release a therapeutic payload. Furthermore, microbubbles can also be targeted to regions of disease by surface conjugation of specific ligands or antibodies, which bind to the disease markers[28]. Recently, genetically encoded gas nanostructure from microorganisms has been demonstrated as a promising candidate as molecular reporters[41]. All these applications imply that microbubbles have high specificity and selectivity, with which TRUME may provide precise optical mediation for drugs or cells or molecules. Example applications range from selective photo-thermal therapy for targeting tumour cells[42], to specific light delivery in optogenetics[43].

## Methods

### Setup

The TRUME experiment was carried out in a custom-built setup. A pulsed laser beam (532 nm wavelength, 7 ns pulse width, 20 kHz repetition rate, 7 mm coherent length) generated from a Q-switch laser (Navigator, Spectra-Physics, USA) was spilt into three beams: a sample beam, a reference beam and a playback beam. Both of the sample beam and the reference beam were shifted by 50 MHz using an acousto-optical modulator (AOM, AFM-502-A1, IntraAction, USA). The interference between the transmitted sample beam and reference beam was measured by the camera (PCO.edge, PCO, Germany) of the DOPC system. The playback beam was modulated with the conjugated phase of the subtracted field by an SLM (Pluto, Holoeye, Germany), which was precisely aligned to the camera through a beam splitter.

The 20 MHz ultrasound burst was generated by a transducer with a 12.7 mm focal length and 6.35 mm element diameter (V317, Olympus, USA), and the 45 MHz ultrasound burst was generated by a transducer that has a 6 mm focal length and 6.35 mm element diameter (nominal frequency at 50 MHz, calibrated peak frequency at 44.4 MHz, V3330, Olympus,

USA). Both transducers were driven by a RF power amplifier (30W1000B, Amplifier Research, USA).

To directly visualize the results, a custom-built microscope with a 20X objective (SLMPlan N, Olympus, Japan) and a tube lens of 100 mm focal length was used to image the target plane to a CCD camera (Stingray F145, Allied Vision Technologies, Germany). To demonstrate the cytometry application, the fluorescent signals were filtered by a 561 nm long-pass (LP02-561RE-25, Semrock, USA) and a 582/75 nm band-pass filter (FF01-582/75-25, Semrock) and detected by a SPCM (SPCM-AQRH-14, Perkinelmer, Canada).

**Signal flow**

The sample beam and reference beam were modulated by 50 MHz signals generated from two channels of a function generator (AFG 3252, Tektronix, USA). The optical field transmitted through the sample was measured by the camera (exposure time: 20 ms, framerate: 35 fps) of the DOPC system using 4-phase shifting based digital holography[34]. The phase shifting was synchronised with the camera exposure by controlling signals from a data acquisition card (DAQ, PCI-6281, NI, USA). The ultrasound burst signal (10 cycles, 10 μs interval) was generated by another function generator (4065, BK Precision, USA) and time-gated (28.6 ms) by the DAQ.

**Sample preparation**

The microbubbles (Optison, GE health care, USA) was diluted to 10% (v/v %) in 1% (w/w%) agarose gel in aqueous phase or 1X PBS (Demonstration of flow stream monitoring) and perfused in an acrylic capillary tube (inner diameter: 50 μm, outer diameter: 100 μm, Paradigm Optics, USA), which was positioned inside a clear polystyrene cuvette. 10% Polyacrylamide gel was used to fill the space in the cuvette to secure the capillary tube. Two diffusers (10 X 10 mm, 220 grit ground glass, Edmund Optics, USA) were placed outside the cuvette in parallel with ~10 mm distance in between. The microbubble sheet was ~20 μm thick and sandwiched between two blocks of agarose gel with dimensions of 10 mm (x) X 10 mm (y) X 3 mm (z). The microbubble sheet was positioned between and parallel to the diffusers. The ultrasound beam was aligned to the microbubble sheet by maximising the amplitude of the echo received from the focus.

In the flow stream monitoring experiment, fluorescent microspheres with 4 μm diameter (FluoSpheres 580/605, Life Science, USA) were used as targets. In the *ex-vivo* tissue experiment, fresh chicken breast tissue was used as diffusive medium. For each tissue diffuser, a piece of 2-mm thick chicken breast tissue slice (10 mm (x) X 10 mm (y)) was sandwiched between two pieces of cover glass separated by a 2-mm spacer.

**Ultrasound beam characterisation**

We calculated theoretical ultrasound pressure field using the fast near field method[44]. We first calculated the pressure fields at different single frequencies ranged from 1 MHz to 100 MHz and summed the profiles with a weight accounting for transducer response and frequency spectrum of ultrasound pulse train.

The ultrasound pressure was measured in room-temperature water using a calibrated hydrophone (HGL-0085, Onda, USA). To characterise the profile of the ultrasound beam, we operated the transducer in pulse-echo mode using a pulser-receiver (5900PR, Olympus, USA) and scanned a line target (air filled polycarbonate tube, inner diameter 22.5 μm, outer diameter 25 μm, Paradigm Optics, USA) by translating the transducer in the lateral and axial direction respectively[45]. This method provides a more accurate measurement than using the hydrophone because the active diameter of the hydrophone is larger than the waist of the ultrasound beam generated by the V3330 transducer. The peak-peak voltages of the echoes were measured by an oscilloscope (DPO 3012, Tektronix, USA). Because the measurement was based on single cycle burst, side lobes were not shown.

**Watershed algorithm**

We first binarized the image with a threshold that was 7 times higher than the background intensity. This step outputs a binary image in which only the pixels around the peak have the value of 1. We then segmented the binary image with a watershed algorithm and extracted the centroid of each focal spots.

## Acknowledgements

The authors thank Dr. Daifa Wang for developing the idea, Mr. Roarke Horstmeyer for the comments on the mathematics and the manuscript, and Mr. Joshua Brake as well as Mr. Edward Zhou for helpful discussions. The author would also like to thank Dr. Mikhail G. Shapiro, Dr. Stephen P. Morgan and Dr. Melissa L. Mather for the discussions on microbubbles. This work is supported by the National Institutes of Health (1DP2OD007307-01), the National Institutes of Health BRAIN Initiative (1U01NS090577-01), and a GIST-Caltech Collaborative Research Proposal (CG2012).

## Author contributions

H.R. and M.J. contributed equally to the work. H.R. conceived the idea. H.R., M.J. and C.Y. developed the idea and designed the experiments, which were carried out by H.R. and M.J. Data analysis was carried out by H.R. and M.J. with help from C.Y., who also supervised the project. All authors contributed to the preparation of the manuscript.

# References

1. Mosk, A. P., Lagendijk, A., Lerosey, G. & Fink, M. Controlling waves in space and time for imaging and focusing in complex media. *Nature Photonics* **6,** 283–292 (2012).

2. Vellekoop, I. M. Feedback-based wavefront shaping. *Opt. Express* **23,** 12189 (2015).

3. Kim, M., Choi, W., Choi, Y., Yoon, C. & Choi, W. Transmission matrix of a scattering medium and its applications in biophotonics. *Opt. Express* **23,** 12648 (2015).

4. Yu, H. *et al.* Recent advances in wavefront shaping techniques for biomedical applications. *Curr. Appl. Phys.* **15,** 632–641 (2015).

5. Vellekoop, I. M. & Mosk, A. P. Focusing coherent light through opaque strongly scattering media. *Opt. Lett.* **32,** 2309–2311 (2007).

6. Yaqoob, Z., Psaltis, D., Feld, M. S. & Yang, C. Optical phase conjugation for turbidity suppression in biological samples. *Nat Phot.* **2,** 110–115 (2008).

7. Cui, M. & Yang, C. Implementation of a digital optical phase conjugation system and its application to study the robustness of turbidity suppression by phase conjugation. *Opt. Express* **18,** 3444–3455 (2010).

8. Popoff, S. M. *et al.* Measuring the Transmission Matrix in Optics: An Approach to the Study and Control of Light Propagation in Disordered Media. *Phys. Rev. Lett.* **104,** 100601 (2010).

9. Yu, H. *et al.* Measuring Large Optical Transmission Matrices of Disordered Media. *Phys. Rev. Lett.* **111,** 153902 (2013).

10. Chaigne, T. *et al.* Controlling light in scattering media non-invasively using the photoacoustic transmission matrix. *Nat. Photonics* **8,** 58–64 (2013).

11. Hsieh, C., Pu, Y., Grange, R. & Psaltis, D. Digital phase conjugation of second harmonic radiation emitted by nanoparticles in turbid media. *Opt. Express* **18,** 533–537 (2010).

12. Vellekoop, I. M., Cui, M. & Yang, C. Digital optical phase conjugation of fluorescence in turbid tissue. *Appl. Phys. Lett.* **101,** 81108 (2012).

13. Katz, O., Small, E., Guan, Y. & Silberberg, Y. Noninvasive nonlinear focusing and imaging through strongly scattering turbid layers. *Optica* **1,** 170 (2014).

14. Zhou, E. H., Ruan, H., Yang, C. & Judkewitz, B. Focusing on moving targets through scattering samples. *Optica* **1,** 227 (2014).

15. Ma, C., Xu, X., Liu, Y. & Wang, L. V. Time-reversed adapted-perturbation (TRAP) optical focusing onto dynamic objects inside scattering media. *Nat. Photonics* **8,** 931–936 (2014).

16. Kong, F. *et al.* Photoacoustic-guided convergence of light through optically diffusive media. *Opt. Lett.* **36,** 2053–5 (2011).

17. Caravaca-Aguirre, A. M. *et al.* High contrast three-dimensional photoacoustic imaging through scattering media by localized optical fluence enhancement. *Opt. Express* **21,** 26671 (2013).

18. Lai, P., Wang, L., Tay, J. W. & Wang, L. V. Photoacoustically guided wavefront shaping for enhanced optical focusing in scattering media. *Nat. Photonics* **9,** 126–132 (2015).

19. Xu, X., Liu, H. & Wang, L. V. Time-reversed ultrasonically encoded optical focusing into scattering media. *Nat. Photonics* **5,** 154–157 (2011).

20. Wang, Y. M., Judkewitz, B., DiMarzio, C. A. & Yang, C. Deep-tissue focal fluorescence imaging with digitally time-reversed ultrasound-encoded light. *Nat Commun* **3,** 928 (2012).

21. Si, K., Fiolka, R. & Cui, M. Fluorescence imaging beyond the ballistic regime by ultrasound pulse guided digital phase conjugation. *Nat. Photonics* **6,** 657–661 (2012).

22. Jang, M., Ruan, H., Judkewitz, B. & Yang, C. Model for estimating the penetration depth limit of the time-reversed ultrasonically encoded optical focusing technique. *Opt Express* **22,** 5787–5807 (2014).

23. Kothapalli, S.-R. & Wang, L. V. Ultrasound-modulated optical microscopy. *J. Biomed. Opt.* **13,** 054046

24. Si, K., Fiolka, R. & Cui, M. Breaking the spatial resolution barrier via iterative sound-light interaction in deep tissue microscopy. *Sci. Rep.* **2,** 748 (2012).

25. Ruan, H., Jang, M., Judkewitz, B. & Yang, C. Iterative time-reversed ultrasonically encoded light focusing in backscattering mode. *Sci. Rep.* **4,** 7156 (2014).

26. Suzuki, Y., Tay, J. W., Yang, Q. & Wang, L. V. Continuous scanning of a time-reversed ultrasonically encoded optical focus by reflection-mode digital phase conjugation. *Opt. Lett.* **39,** 3441–4 (2014).

27. Judkewitz, B., Wang, Y. M., Horstmeyer, R., Mathy, A. & Yang, C. Speckle-scale focusing in the diffusive regime with time-reversal of variance-encoded light (TROVE). *Nat. Photonics* **7,** 300–305 (2013).

28. Lindner, J. R. Microbubbles in medical imaging : current applications and future directions. **3,** 527–532 (2004).

29. Goertz, D. E. *et al.* High frequency nonlinear B-scan imaging of microbubble contrast agents. *IEEE Trans. Ultrason. Ferroelectr. Freq. Control* **52,** 65–79 (2005).

30. Benchimol, M. J. *et al.* Phospholipid/Carbocyanine Dye-Shelled Microbubbles as Ultrasound-Modulated Fluorescent Contrast Agents. *Soft Matter* **9,** 2384–2388 (2013).

31. Liu, Y., Feshitan, J. A., Wei, M.-Y., Borden, M. A. & Yuan, B. Ultrasound-modulated fluorescence based on fluorescent microbubbles. *J. Biomed. Opt.* **19,** 085005 (2014).

32. Ruan, H., Mather, M. L. & Morgan, S. P. Ultrasound modulated optical tomography contrast enhancement with non-linear oscillation of microbubbles. *Quant. Imaging Med. Surg.* **5,** 9–16 (2015).

33. Jang, M., Ruan, H., Zhou, H., Judkewitz, B. & Yang, C. Method for auto-alignment of digital optical phase conjugation systems based on digital propagation. *Opt. Express* **22,** 14054–71 (2014).

34. Yamaguchi, I., Matsumura, T. & Kato, J.-I. Phase-shifting color digital holography. *Opt. Lett.* **27,** 1108–10 (2002).

35. Shekhar, H., Rychak, J. J. & Doyley, M. M. Modifying the size distribution of microbubble contrast agents for high-frequency subharmonic imaging. *Med. Phys.* **40,** 082903 (2013).

36. Pancholi, K. P., Farook, U., Moaleji, R., Stride, E. & Edirisinghe, M. J. Novel methods for preparing phospholipid coated microbubbles. *Eur. Biophys. J.* **37,** 515–20 (2008).

37. Palanchon, P., Klein, J. & de Jong, N. Production of standardized air bubbles: Application to embolism studies. *Rev. Sci. Instrum.* **74,** 2558 (2003).

38. Jang, M. *et al.* Relation between speckle decorrelation and optical phase conjugation (OPC)-based turbidity suppression through dynamic scattering media: a study on in vivo mouse skin. *Biomed. Opt. Express* **6,** 72 (2015).

39. Chomas, J. E., Dayton, P., Allen, J., Morgan, K. & Ferrara, K. W. Mechanisms of contrast agent destruction. *IEEE Trans. Ultrason. Ferroelectr. Freq. Control* **48,** 232–48 (2001).

40. Ferrara, K., Pollard, R. & Borden, M. Ultrasound microbubble contrast agents: fundamentals and application to gene and drug delivery. *Annu. Rev. Biomed. Eng.* **9,** 415–47 (2007).

41. Shapiro, M. G. *et al.* Biogenic gas nanostructures as ultrasonic molecular reporters. *Nat. Nanotechnol.* **9,** 311–6 (2014).

42. El-Sayed, I. H., Huang, X. & El-Sayed, M. A. Selective laser photo-thermal therapy of epithelial carcinoma using anti-EGFR antibody conjugated gold nanoparticles. *Cancer Lett.* **239,** 129–35 (2006).

43. Gradinaru, V., Mogri, M., Thompson, K. R., Henderson, J. M. & Deisseroth, K. Optical deconstruction of parkinsonian neural circuitry. *Science* **324,** 354–9 (2009).

44. Chen, D. & McGough, R. J. A 2D fast near-field method for calculating near-field pressures generated by apodized rectangular pistons. *J. Acoust. Soc. Am.* **124,** 1526–37 (2008).

45. Raum, K. & O'Brien, W. D. Pulse-echo field distribution measurement technique for high-frequency ultrasound sources. *IEEE Trans. Ultrason. Ferroelectr. Freq. Control* **44,** 810–815 (1997).